\documentclass[conference]{IEEEtran}
\IEEEoverridecommandlockouts
\usepackage{cite}
\usepackage{amsmath,amssymb,amsfonts}
\usepackage{algpseudocode}
\usepackage{graphicx}
\usepackage{textcomp}
\usepackage{xcolor}

\DeclareMathOperator*{\argmax}{arg\,max}
\def\BibTeX{{\rm B\kern-.05em{\sc i\kern-.025em b}\kern-.08em
    T\kern-.1667em\lower.7ex\hbox{E}\kern-.125emX}}
\begin{document}

\title{Reinforcement Learning Quantum Local Search\\
}

\author{
\IEEEauthorblockN{
    Chen-Yu Liu \IEEEauthorrefmark{1}\IEEEauthorrefmark{3}, and Hsi-Sheng Goan \IEEEauthorrefmark{2}
}
\IEEEauthorblockA{\IEEEauthorrefmark{1} Graduate Institute of Applied Physics, National Taiwan University, Taipei, Taiwan}
\IEEEauthorblockA{\IEEEauthorrefmark{2} Department of Physics, National Taiwan University, Taipei, Taiwan}

\IEEEauthorblockA{Email:\IEEEauthorrefmark{3}  d10245003@g.ntu.edu.tw}

}

\maketitle

\begin{abstract}
Quantum Local Search (QLS) is a promising approach that employs small-scale quantum computers to tackle large combinatorial optimization problems through local search on quantum hardware, starting from an initial point. However, the random selection of the sub-problem to solve in QLS may not be efficient. In this study, we propose a reinforcement learning (RL) based approach to train an agent for improved sub-problem selection in QLS, beyond random selection. Our results demonstrate that the RL agent effectively enhances the average approximation ratio of QLS on fully-connected random Ising problems, indicating the potential of combining RL techniques with Noisy Intermediate-scale Quantum (NISQ) algorithms. This research opens a promising direction for integrating RL into quantum computing to enhance the performance of optimization tasks.
\end{abstract}

\begin{IEEEkeywords}
Reinforcement Learning, Quantum Optimization, Local Search, 
\end{IEEEkeywords}

\section{Introduction}

Exploiting the principles of quantum mechanics, quantum computing facilitates calculations and problem-solving with unparalleled efficiency \cite{qc1, qc2}. This technology holds immense promise for expediting solutions to specific problems beyond the capabilities of classical computers, thereby offering a myriad of application opportunities. To solve problems on quantum computers, it is necessary to map the original problem to a form that can be addressed by quantum algorithms. One common approach for tackling a wide variety of combinatorial optimization problems is to convert them into quadratic unconstrained binary optimization (QUBO) problems, map the problem to an Ising Hamiltonian, and then determine the Hamiltonian's ground state by variational quantum eigensolver (VQE) \cite{vqe0, vqe1, vqe2, vqe3, vqe4} or quantum annealing \cite{toising1, portfolio2}, where the ground state corresponds to the solution of the original problem \cite{toising1, toising2}. Quantum optimization has been successfully applied to real-world challenges, such as portfolio optimization \cite{portfolio2, portfolio3, portfolio4, portfolio5}, industrial optimization \cite{supplychain1, industry1}, recommendation system \cite{qrs1, fsqc1}, and traveling salesman problems \cite{tsp1, tsp2}.

Given the limitations on the number of qubits during the Noisy Intermediate-scale Quantum (NISQ) era, several methods have been proposed to decrease the number of qubits required for VQE and/or quantum annealing, including cluster-based approaches, large-system sampling approximation, quantum local search (QLS), and other techniques \cite{qls1 ,dcqaoa1, clustervqe1, qaoainqaoa1, qbeff1, hgaip, qbsolv1}. 

The QLS algorithm is designed to address combinatorial optimization problems by iteratively solving sub-problems and updating local configurations, such that the required number of qubits is just the size of the sub-problem instead of the full problem. This approach enables the use of both gate-based quantum chips and quantum annealers for solving sub-problems. The random selection of sub-problems offers opportunities for enhancement by investigating alternative sub-problem selection strategies, such as the reinforcement learning (RL) strategy explored in this work.

\section{Preliminary}
In this preliminary section, we review the key ingredients of our work: RL and QLS. RL is a computational approach that enables agents to learn optimal decision-making strategies through interaction with their environment, while QLS is a technique that leverages small quantum computers to solve large combinatorial optimization problems by performing local search on quantum hardware.

\begin{figure*}
\centering
    \includegraphics[scale=0.33]{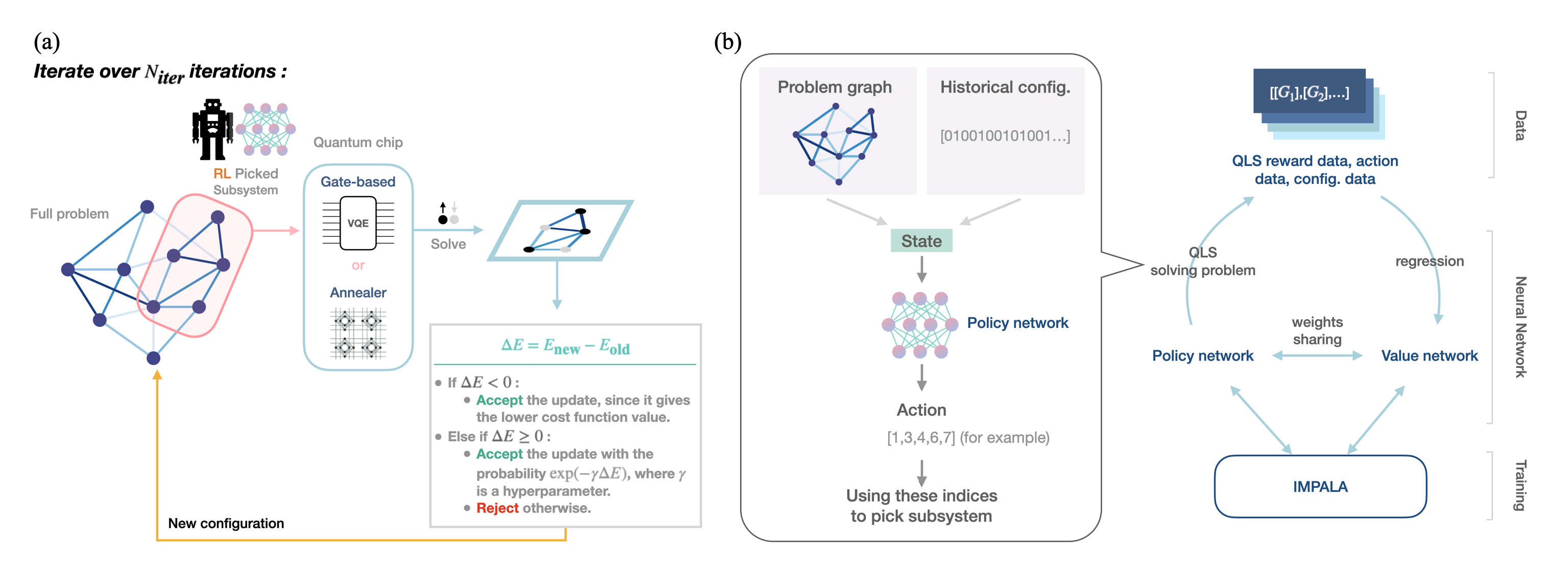}
    \caption{(a) In the RL-QLS scheme, with $N_{\text{iter}}$ iterations, the RL agent selects the sub-problem to be solved by either a gate-based quantum chip or a quantum annealer from the full problem and accepts the trial solution update under specific conditions. (b) The training scheme for RL-QLS involves the policy network/agent determining the sub-problem based on the problem graph information and the historical configuration of trial solutions. It generates data for different problem graphs $G_i$ to train the neural networks using the IMPALA algorithm.
    } 
\label{fig:flow}
\end{figure*}

\subsection{Quantum Local Search}

Quantum Local Search (QLS) \cite{qls1, qbsolv1} combines the principles of quantum computing with local search techniques to solve combinatorial optimization problems. In combinatorial optimization, the goal is to find the optimal solution within a large, discrete solution space, often consisting of various possible configurations. Classical local search \cite{cls1} algorithms operate by iteratively exploring neighboring solutions in the solution space and moving towards better solutions based on a predefined criterion. QLS utilizes small-scale quantum computers to perform local search operations and leverage the power of quantum computing to enhance the search process. By employing quantum computing, QLS can explore multiple neighboring solutions simultaneously due to quantum parallelism, leading to a potentially more efficient search process.

For an Ising problem with $N$ variables that corresponding to some combinatorial optimization problems: 
\begin{equation}
    H = \sum_{i,j = 1}^N J_{ij} \sigma_i^z \sigma_j^z + \sum_{i=1}^N h_i \sigma_i^z,
    \label{ising}
\end{equation}
with $J_{ij}$ the coupling terms and $h_i$ the linear terms. 
Given an initial starting point of the trial solution state $| \psi \rangle$, QLS iteratively and randomly selects sub-problems with size $m$ such that: 
\begin{equation}
    H_{\text{sub}} = \sum_{i',j' = 1}^m J_{i'j'} \sigma_{i'}^z \sigma_{j'}^z + \sum_{i'=1}^m h_{i'} \sigma_{i'}^z,
\end{equation}
within the larger optimization problem $H$ and solves them using quantum algorithms such as VQE on gate-based quantum computer or quantum annealing on quantum annealer. The algorithm then updates the current solution $| \psi \rangle$ of size $N$ based on the results $| \phi \rangle$ of size $m$ obtained from the quantum computation, moving towards an improved solution. This process is repeated until a satisfactory solution is found or a predefined stopping criterion is met.

The random selection process of subsystems in QLS presents potential inefficiencies in solution exploration. These inefficiencies include redundancy, imbalance, and lack of direction. Redundancy arises when the algorithm chooses similar or already explored sub-problems, wasting computational resources and time. Imbalance occurs when the random selection fails to explore the solution space evenly, possibly overlooking better solutions. The lack of direction results from not utilizing problem-specific knowledge or learning from past iterations to guide the selection process.

To address these issues, alternative approaches, such as reinforcement learning (RL) agents, can be employed to enhance sub-problem selection in QLS. This adaptive approach can improve solution exploration efficiency, enabling the algorithm to discover superior solutions more effectively.


\subsection{Reinforcement Learning}

RL \cite{rl1} is a subfield of machine learning that focuses on training agents to make decisions by interacting with an environment. In RL, an agent learns to perform actions that maximize cumulative rewards over time. The learning process is typically modeled as a Markov Decision Process (MDP), which consists of state space $\mathbb{S}$, action space $\mathbb{A}$, reward function, and transition probabilities. The goal is to learn an optimal policy $\pi^*(a|s)$, which dictates the best action $a \in \mathbb{A}$ to take in each state $s \in \mathbb{A}$, 
that maximizes the expected discounted return:
\begin{equation}
    \pi^* = \argmax_{\pi} \mathbb{E}_{\pi} \left[ \sum_{t=0}^{T} \Gamma^t R_t \right],
\end{equation}
where $R_t$ is the reward obtained at time step $t$, and $\Gamma \in [0, 1]$ is a discount factor that determines the relative importance of immediate versus future rewards. 

In this work, we use Importance Weighted Actor-Learner Architecture (IMPALA) \cite{impala1} to train the RL agent. IMPALA is a distributed deep RL algorithm designed to scale up the learning process using multiple parallel actor-learner processes. It achieves this by decoupling the acting and learning processes, allowing for efficient use of computational resources. The actors generate trajectories $(s_t, a_t, R_t, s_{t+1})$ by interacting with the environment using the current policy, while the learner updates the policy parameters $\theta$ using off-policy learning with a value function critic $V(s; \theta_v)$. The policy gradient update is performed using the loss function:
\begin{equation}
    \mathcal{L}(\theta) = \mathbb{E}{\tau} \left[\sum_{t=0}^{T} \rho_t \nabla_\theta \log \pi_\theta(a_t | s_t) \delta_t^{\pi_\theta}\right], 
\end{equation}
where $\tau$ represents a trajectory generated by policy, $\rho_t = \frac{\pi_\theta(a_t|s_t)}{\mu_\theta(a_t|s_t)}$ is the importance sampling ratio with $\mu_\theta$ the older policy which generates trajectories in replay buffer, and $\delta_t^{\pi_\theta} = R_t + \Gamma V(s_{t+1}; \theta_v) - V(s_t; \theta_v)$ is the temporal difference (TD) error.

The use of importance sampling in IMPALA allows for stable and efficient off-policy learning, mitigating the issues of stale trajectories generated by the actors. The distributed nature of the algorithm enables the handling of large-scale environments and high-dimensional state spaces. By combining the advantages of distributed RL with off-policy learning, IMPALA offers a scalable and powerful solution for a wide range of reinforcement learning problems.

\section{Reinforcement Learning Quantum Local Search}

\begin{figure*}
\centering
    \includegraphics[scale=0.30]{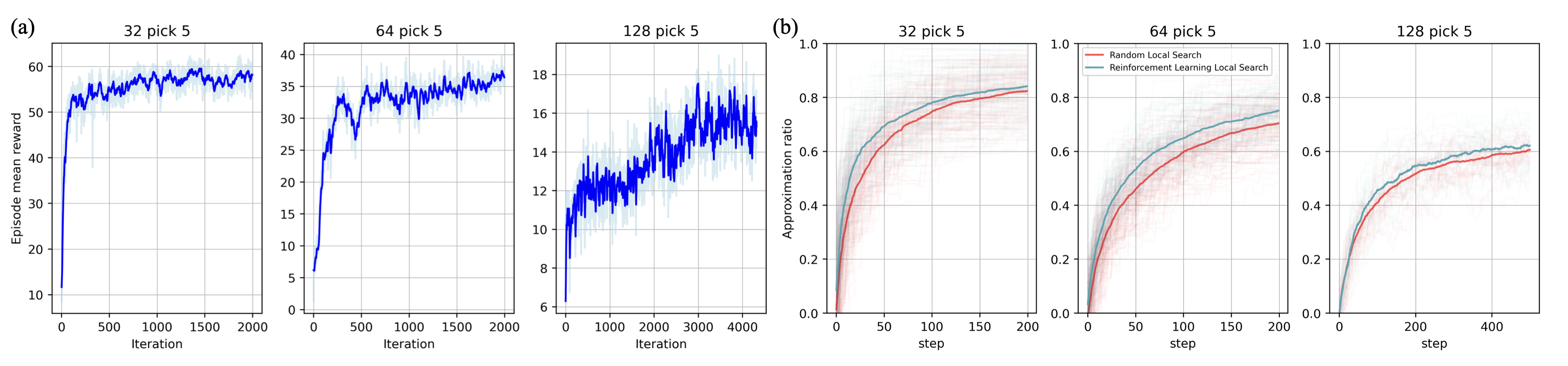}
    \caption{(a) Episode mean reward during the training process, where ``$N$ pick $m$" labels represent the size-$N$ full problem and corresponding size-$m$ sub-problem. (b) The approximation ratio results of RL-QLS (blue) applied to the Ising problems testing set compared with that of random sub-problem selection strategy (red). The solid line represents the average value among the $N_t$ tests, while individual test results are shown as faded lines. For the ``32 pick 5" and ``64 pick 5" cases, $N_t = 100$, while for the ``128 pick 5" case, $N_t = 30$.
    } 
\label{fig:result}
\end{figure*}

In this work, we propose Reinforcement Learning Quantum Local Search (RL-QLS) to replace the random selection process of sub-problems in QLS with a strategy represented by a trained RL agent, as depicted in the schematic diagram in Fig.~\ref{fig:flow}(a). To train such an agent within the QLS framework, the formulation of fundamental elements of RL, including state, action, reward, and environment, is required. The agent's state, $s = (G, \vec{\phi})$, is constructed using information from the size-$N$ problem graph $G = (J_{11}, J_{12}, \hdots, J_{NN}, h_1, \hdots, h_N)$, where $J_{ij}$ represents coupling terms and $h_i$ represents linear terms as in \eqref{ising}, along with the historical configuration $\vec{\phi}$ of the trial solution up to the previous 1 step. The agent's action $a = (q_1, q_2, \hdots, q_m)$ is a list consisting of $m$ elements, where the elements $q_i \in \{0, 1, \hdots, N-1\}$ represent the indices of the variables for the original problem of size $N$ and $m$ is the size of the sub-problem. The reward $\mathbf{R}$ is defined as the approximation ratio $R_\textrm{ar} = \textrm{TSE}/\textsf{Dwave-tabu }\textrm{GSE}$, where TSE is the trial solution energy and \textsf{Dwave-tabu} GSE is the ground state energy obtained by the \textsf{Dwave-tabu} solver. The environment dynamics for the RL agent involve a flow from inputting the action to outputting the reward and the subsequent state. In the context of QLS, one step of the environment dynamics entails the following sub-steps:

\begin{enumerate}
\item For a state $s = (G, \vec{\phi})$, policy $\pi(a|s)$ produces an action $a$.
\item For the sub-problem selected by the action $a$, compute the sub-problem solution and calculate the corresponding new trial solution energy $E_{\text{new}}$ for the full problem.
\item With $E_{\text{old}}$ representing the old trial solution energy, accept the update if $\Delta E = E_{\text{new}} - E_{\text{old}} < 0$. If $\Delta E \ge 0$, the update is accepted with probability $e^{-\gamma \Delta E}$; otherwise, the update is rejected. Here, $\gamma$ is a hyperparameter.
\item Calculate the reward for the new configuration (trial solution).
\item Construct the new state using the above information, $s_{\text{new}} = (G, \vec{\phi}_{\text{new}})$, while the graph information remains unchanged.
\end{enumerate}

\subsection{Training the agent}
Following the description of the basic elements of RL in the QLS scheme, the training process for the RL agent using the IMPALA algorithm in this work is set up as follows: For each training iteration, 100 episodes are evaluated, with each episode consisting of 200 steps of quantum local search (environment dynamics) iterations to generate reward data. A set of 1000 randomly generated fully-connected Ising problems serve as the training problem set, such that for each episode, the agent will randomly start from one of these problems. This configuration enables the RL agent to gain a more general understanding of solving various Ising problems. The schematic diagram of the RL training is shown in Fig.~\ref{fig:flow}(b). 

The training problem set consists of Ising problems defined as in \eqref{ising}, the range of the couplings and linear terms are $J_{ij} \in (-1, 1)$ and $h_i \in (-1, 1)$. The hyperparameter $\gamma$, used to determine the acceptance rate of updates, is set to $\frac{100}{\textsf{Dwave-tabu } \textrm{GSE}}$, so that the energy difference is normalized for different energy scale. Three cases of size-$N$ Ising problems are examined, where $N \in \{ 32, 64, 128 \}$. The sub-problem size $m$ is set to 5 for all three cases of size-$N$ Ising problems, so that quantum solvers with 5 qubits can be employed. The \textsf{Dwave-tabu} solver is utilized as a quantum simulator for solving the sub-problems to assess the effectiveness of the RL sub-problem selection strategy. 

The episode mean reward during the training process is depicted in Fig.~\ref{fig:result}(a), with the ``$N$ pick $m$" labels representing the size-$N$ full problem and corresponding size-$m$ sub-problem. For the ``32 pick 5" case, 2000 training iterations are utilized, and the episode mean reward converges to approximately 60 by the end of the training process. In the ``64 pick 5" case, 2000 training iterations are also used, exhibiting similar behavior to the ``32 pick 5" case, but with an episode mean reward of about 37 at the end. The ``128 pick 5" case employs 4300 training iterations, and the episode mean reward approaches 16 by the end of the training process. The observation that cases with larger $N$ values have smaller overall episode mean rewards reflects the increased difficulty associated with solving larger-sized problems.

\subsection{Testing the agent}

To evaluate the trained agent, three testing problem sets corresponding to the three cases of problem size $N$ are generated, with couplings and linear terms in the same range as in the training problem set. For the cases ``32 pick 5'' and ``64 pick 5'', $N_t = 100$ test problems are assessed, while for the case ``128 pick 5'', $N_t = 30$ test problems are examined. The test results are displayed in Fig.~\ref{fig:result}(b), with the results of RL-QLS depicted in red and those of the random sub-problem selection strategy shown in blue for comparison. For the cases ``32 pick 5'' and ``64 pick 5'', optimization is performed for 200 steps, and 500 steps are used for the case ``128 pick 5''. A noticeable improvement can be observed for the RL-QLS compared to the original random sub-problem selection strategy.

\section{Discussion}

The training and testing processes revealed that the RL agent's performance was affected by the problem size, with smaller episode mean rewards observed for larger problems. This result highlights the increased difficulty of solving larger-sized problems and suggests that further research may be needed to optimize the agent's performance for these more challenging tasks. The RL-QLS method has shown promise in addressing QLS problems and improving sub-problem selection strategies.

The potential ways to improve the performance of the RL agent could be refining the state representation to incorporate more complex problem information or richer historical configurations, and the exploration of alternative RL algorithms and training techniques could potentially enhance the agent's performance and adaptability, enabling it to tackle more complex and diverse problem sets. 

Finally, conducting experiments on real quantum devices would provide a valuable assessment of the practical applicability of the RL-QLS approach in real-world scenarios, highlighting areas for further exploration and enhancement.

\section{Conclusion}

In this work, we introduced RL-QLS, an approach that leverages RL to enhance the sub-problem selection process in QLS. The RL-QLS method replaces the random selection process in QLS with a trained RL agent, which adapts its strategy based on its interaction with the environment. This adaptive approach enables the RL-QLS algorithm to improve the exploration efficiency of combinatorial optimization problems, leading to more effective discovery of superior solutions. We implemented the IMPALA algorithm to train the RL agent, making it capable of handling large-scale environments and high-dimensional state spaces.

The simulation results demonstrated that the RL-QLS algorithm outperformed the random sub-problem selection strategy in solving Ising problems (combinatorial optimization problems), showing the effectiveness of the RL approach in enhancing the QLS method. The episode mean rewards obtained during the training process indicated that the RL agent was able to learn from the environment and improve its performance over time. In the testing phase, the RL-QLS algorithm achieved better approximation ratios compared to the random selection strategy, validating the potential of the proposed method in solving large-scale optimization problems.

In conclusion, the RL-QLS algorithm presents a promising direction for solving problems in combinatorial optimization (Ising problem) by exploiting the capabilities of quantum computing and reinforcement learning. Future work can apply the RL-QLS approach to other optimization problems and examine the scalability of the algorithm as quantum computing resources improve. Integrating more advanced RL techniques and investigating the transferability of the trained agent can further enhance the performance and efficiency of the RL-QLS method, providing even better solutions to complex optimization challenges.



\end{document}